# Accurate Calculation of Switching Events in Electromagnetic Transient Simulation Considering State Variable Discontinuities

Sheng Lei
Hitachi Energy
Santa Clara, CA, USA
sheng.lei@hitachienergy.com

*Abstract*—Accurate calculation of switching events is important for electromagnetic transient simulation to obtain reliable results. The common presumption of continuous differential state variables could prevent the accurate calculation, thus leading to unreliable results. This paper explores accurately calculating switching events without presuming continuous differential state variables. Possibility of the calculation is revealed by the proposal of related methods. Feasibility and accuracy of the proposed methods are demonstrated and analyzed via numerical case studies.

## I. INTRODUCTION

Electromagnetic transient (EMT) simulation has been widely applied to study the dynamic behaviors of power electronics-integrated power systems where switching events frequently occur, which usually appear in the form of fault application/removal, tripping/connecting of a device, turning on/off of a power electronic valve, hitting a limit in a controller, etc.. Accurate calculation of switching events is a significant factor that can greatly impact the fidelity and reliability of the simulation results, thus attracting a lot of research interest and effort [1]-[6].

In power system EMT simulation, device-level or system-level dynamics are the study focus; detailed valve-level transients are not of interest. Therefore, power electronic valves are generally modeled by idealized switches [2], [4], [6]-[8]. Classical power system EMT simulators typically use the implicit trapezoidal method as the main numerical integrator with a user-defined fixed step size to discretize the differential equations of the studied systems [9]-[10]. Some simulators use other numerical integrators, such as TR-BDF2 [8]. Mathematical models of network elements (e.g., RLC branches, transformers) are converted into Norton equivalent circuits and organized via Nodal Analysis [9]-[10].

When calculating switching events, references in the literature tend to presume continuity of differential state variables (e.g., inductor currents, capacitor voltages) [4]-[6], [8]. Such presumption actually prevents Dirac impulses. Logically, a Dirac impulse is the derivative of a step function, which mathematically characterizes an instantaneous change. Under the presumption, step functions are not allowed for differential state variables, so that Dirac impulses are not possible. Nevertheless, Dirac impulses can appear if idealized switches are adopted [2], [11]-[12]. Furthermore, potential Dirac impulses indicate necessary status changes of switches, which are indispensable in resolution of simultaneous switching [2], [12]. Resolution of simultaneous switching will be discussed in detail in Section III-D of this paper.

To the author's best knowledge, only few references explicitly allow discontinuous differential state variables in power system EMT simulation. Reference [2] repeatedly performs a reduced time step of backward Euler calculation to resolve simultaneous switching, and then performs another one to reinitialize the simulation run, taking inspiration from the ZZ model method for circuit simulation [11]. However, this method is less accurate, as will be discussed in Section III of this paper. It is an interesting observation that continuity of differential state variables is rarely presumed in circuit simulation [11]-[12].

From an engineering point of view, it is reasonable to adopt simplifying presumptions and less accurate methods so as to achieve efficiency. Nevertheless, accurate methods are still needed under certain circumstances where high fidelity results are of interest or required. This paper is to explore accurate calculation of switching events under the classical off-line power system EMT simulation framework without the common presumption of continuous differential state variables, following the spirit of [2] and further developing related methods. Main contributions of the paper are twofold. First, possibility of accurately calculating switching events is revealed without presuming continuous differential state variables, by the proposal of corresponding methods. Second, feasibility and accuracy order of the proposed methods are analyzed via numerical case studies.

The rest of this paper is organized as follows. Section II elucidates the mathematical model for EMT simulation, state variable waveforms around switching events, and tasks involved in calculation of switching events. Section III details aspects of the calculation and proposes related methods. Feasibility and accuracy of the methods are analyzed via numerical case studies in Section IV. Finally, Section V concludes the paper and points out some directions for future research.

## II. POWER SYSTEM MODEL AND STATE VARIABLE WAVEFORMS

Given the statuses of its switches, a generic power system can be modeled as the following nonlinear differential-algebraic equation set

$$\begin{cases} \underline{\dot{x}} = \underline{f}_{\underline{\xi}}(t, \underline{x}, \underline{y}) \\ 0 = \underline{g}_{\underline{\xi}}(t, \underline{x}, \underline{y}) \end{cases} \quad (1)$$

where $t$ denotes the time; $\underline{x}$ denotes the differential state variable; $\underline{y}$ denotes the algebraic state variable; $\underline{\xi}$ denotes the statuses of the switches; $f$ and $g$ are functions depending on $\underline{\xi}$. Except for $t$, all the quantities are vector-valued.

Suppose that a set of switching events occurs at the time instant $t_{SW}$, and thus the switch statuses change, say, $\underline{\xi}[n-1]$ changes to $\underline{\xi}[n]$, where $n$ denotes the index of the status set. Fig. 1 demonstrates waveforms of a generic entry $x$ in $\underline{x}$ and its derivative $\dot{x}$. It is possible that $x$ experiences an instantaneous change. Correspondingly, $\dot{x}$ exhibits a Dirac impulse. In this situation, three sets of values are recognized, which are at $t_{SW-}$, $t_{SW}$ and $t_{SW+}$ respectively. Generally speaking, the following relations hold

$$\begin{aligned} x(t_{SW-}) &\neq x(t_{SW}) = x(t_{SW+}) \\ \dot{x}(t_{SW-}) &\neq \dot{x}(t_{SW}) \neq \dot{x}(t_{SW+}) \end{aligned} \quad (2)$$

In the literature, the calculation for $t_{SW-}$ is sophisticated, which is typically done by linear interpolation [1]-[6] or quadratic interpolation [8]. Nevertheless, most references further presume continuous differential state variables [4]-[6], [8]. Taking (2) for example, it is further presumed that $x(t_{sw-}) = x(t_{sw})$. Consequently, only algebraic state variables can experience instantaneous changes; $t_{SW}$ and $t_{SW+}$ are not distinguished.

This paper, however, explores accurate calculation for $t_{SW}$ and $t_{SW+}$, the combination of which is understood as calculation of switching events in this paper. It does not presume continuity of differential state variables. Some methods for the calculation will be proposed in the coming section.

## III. ACCURATE CALCULATION OF SWITCHING EVENTS CONSIDERING STATE VARIABLE DISCONTINUITIES

Calculation of switching events consists of two phases: resolution of simultaneous switching (for $t_{SW}$) and reinitialization (for $t_{SW+}$). Some methods can be applied to these phases respectively, forming the building blocks of the methods for calculation of switching events.

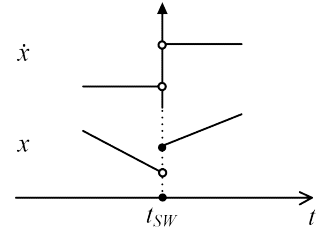

Figure 1. Waveforms of a generic differential state variable and its derivative around the time instant of a swtiching event.

### A. Using a Half Time Step of the Backward Euler Method

Reference [2] calculates the values at a time step after $t_{SW}$ to determine the switch statuses. The step size can be a half of the original step size $h$ of the simulation run [2], which is commonly used in EMT simulation for dealing with switching events. In fact, the values at $t_{SW}+h/2$ are considered as an approximation to those at $t_{SW}$ with this method [11]. The same method is used in circuit simulation as well [12].

"Instantaneous Solution" is another method in the literature for calculating the values at $t_{SW}$ [3]. Nevertheless, [6] reports the "one time step advance error" of the method and also suggests calculating $t_{SW}+h/2$ using the backward Euler method instead.

Specifically, taking the $x$ and $\dot{x}$ in Section II for example, $t_{SW}+h/2$ is calculated with the backward Euler method as

$$x(t_{SW} + \frac{h}{2}) = x(t_{SW-}) + \frac{h}{2}\dot{x}(t_{SW} + \frac{h}{2}) \quad (3)$$

And the values at $t_{SW}$ are approximated by

$$\begin{cases} x(t_{SW}) \approx x(t_{SW} + \frac{h}{2}) \\ \dot{x}(t_{SW}) \approx \dot{x}(t_{SW} + \frac{h}{2}) \end{cases} \quad (4)$$

Apparently, this approximation is only good for constants and generally inaccurate. Switch status checking based on the values at $t_{SW}+h/2$ will incorrectly force all the switching events between $t_{SW}$ and $t_{SW}+h/2$ to take place at $t_{SW}$, leading to unreliable results.

### B. Using Two Half Time Steps of the Backward Euler Method and an Intermediate Linear Extrapolation

This paper puts forward a novel method for calculating the values at $t_{SW}$, using two half time steps of the backward Euler method and an intermediate linear extrapolation. The method is inspired by [4] but without the presumption of continuous differential state variables. Specifically, $t_{SW}+h/2$ is calculated with the backward Euler method as (3). $t_{SW} - h/2$ is calculated with linear extrapolation as

$$x(t_{SW} - \frac{h}{2}) = 2x(t_{SW-}) - x(t_{SW} + \frac{h}{2}) \quad (5)$$

And then $t_{SW}$ is calculated with the backward Euler method as

$$x(t_{SW}) = x(t_{SW} - \frac{h}{2}) + \frac{h}{2}\dot{x}(t_{SW}) \quad (6)$$

## C. Using Two Consecutive Half Time Steps of the Backward Euler Method and a Linear Extrapolation

This paper also proposes another novel method for calculating the values at $t_{SW}$, using two consecutive half time steps of the backward Euler method and a linear extrapolation. The method is inspired by a method reported in [5]. However, continuous differential state variables are again not presumed here. In this method, $t_{SW}+h/2$ is still calculated with the backward Euler method as (3). $t_{SW}+h$ is calculated with the backward Euler method as

$$x(t_{SW}+h) = x(t_{SW}+\frac{h}{2}) + \frac{h}{2}\dot{x}(t_{SW}+h) \quad (7)$$

And then $t_{SW}$ is calculated with linear extrapolation as

$$\begin{cases} x(t_{SW}) = 2x(t_{SW}+\frac{h}{2}) - x(t_{SW}+h) \\ \dot{x}(t_{SW}) = 2\dot{x}(t_{SW}+\frac{h}{2}) - \dot{x}(t_{SW}+h) \end{cases} \quad (8)$$

## D. Resolution of Simultaneous Switching

Simultaneous switching refers to the phenomena in which one or more switching events trigger a series of others at the same time. It should be emphasized that simultaneous switching is not multiple switching events that happen to occur within the same time step. Instead, the switching events occur at the same time instant and are causally related. Examples of simultaneous switching can be found in the literature [1]-[2], [4], [12]-[13].

Resolution of simultaneous switching is to determine the final switch statuses and variable values at $t_{SW}$, which involves repetitive execution of the aforementioned three building blocks, as well as switch status checking and changing. In particular, the following algorithm is implemented:
1) Starting from the values at $t_{SW-}$ and the previously determined switch statuses, calculate the values at $t_{SW}$ using any one of the building blocks. During this calculation, the switch statuses are kept unchanged.
2) Check the status of all the switches according to the values at $t_{SW}$. Change the switch statuses if needed.
3) If one or more switches should change status, discard the interim values at $t_{SW}$, go back to Step 1 and perform a new round of calculation. If no switch should change status, the final switch statuses and variable values at $t_{SW}$ are obtained. Variable values at $t_{SW}$ could exhibit Dirac impulses if necessary.

## E. Reinitialization

Reinitialization is to calculate the values at $t_{SW+}$ so that the simulation run can be resumed. The calculation should be sufficiently accurate because the values, combined with those at $t_{SW}+h$, are utilized to determine and locate the switching events within the time step between these two time instants, if any. The switch statuses during reinitialization, on the other hand, are results from resolution of simultaneous switching and remain unchanged.

The aforementioned three building blocks can be applied to reinitialization with only tiny modifications: in (3)-(8), $t_{SW}$ is changed to $t_{SW+}$ while $t_{SW-}$ is changed to $t_{SW}$. After simultaneous switching has been resolved, any one of the building blocks is executed for reinitialization. Unlike resolution of simultaneous switching, the selected building block only needs to be executed once.

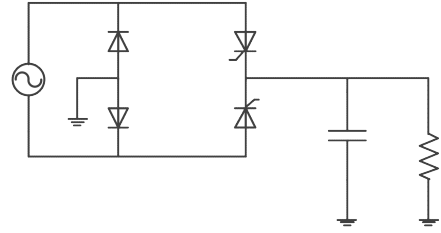

Figure 2. Single-phase semi-controlled full bridge rectifier driving a resistance load.

## F. Methods for Calculation of Switching Events

This paper is going to investigate three methods for calculation of switching events in the next section, though other combinations of the building blocks are also possible. Method A is based on the backward Euler method. Method B is based on the building block in Section III-B. Method C is based on the building block in Section III-C.

Note that Method A is not identical to the method in [2], which calculates $t_{SW}+h$ after resolution of simultaneous switching and resumes the simulation run. With that method, switching events between $t_{SW}+h/2$ and $t_{SW}+h$ are not checked and will thus be missed, further leading to unreliable results. On the other hand, Method A calculates $t_{SW+}$ after resolution of simultaneous switching, learning from [11] and [12].

## IV. NUMERICAL CASE STUDIES

The methods for calculation of switching events have been integrated into a typical time stepping scheme for EMT simulation implemented in MATLAB, which uses the implicit trapezoidal method with a user-defined fixed step size to carry out simulation runs, and linear interpolation to locate and bring the system back to the time instants of switching events. The whole studied system (including network, electric machines, controllers, etc.) is solved simultaneously without artificial time delays, which is necessary for accurate calculation of switching events [2]. Details of the time stepping scheme can be found in [13]. The methods can be easily embedded into the classical EMT simulation framework, with only minor updates in the time loop. The updated scheme allows a user to choose any one of the three methods.

Results from the time stepping scheme adopting the three methods will be compared to those from Simulink applied to the same test systems. Feasibility of the methods can be examined through the comparison. All the Simulink simulations adopt the variable-step solver ode23tb with a maximum step size of 1 μs and a relative tolerance of $10^{-5}$. A variable-step solver is used as the benchmark because it is generally considered more suitable and accurate for treating switching events [4], [14]. Quantities in this section will be expressed in per-unit values unless otherwise stated.

### A. Single-Phase Semi-Controlled Full Bridge Rectifier

Fig. 2 shows a single-phase semi-controlled full bridge rectifier driving a resistance load. The magnitude, phase angle and frequency of the AC voltage source are 1.0, 0 rad

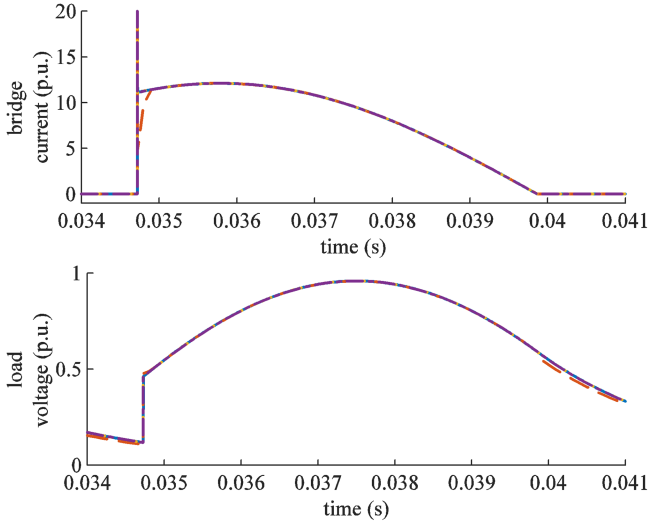

Figure 3. Results from the single-phase semi-controlled full bridge rectifier driving a resistance load. Solid line: Simulink. Dashed line: Method A. Dotted line: Method B. Dash-dotted line: Method C. Spikes and jumps are precisely represented as vertical lines rather than sharp slopes. The dotted line and dash-dotted line basically overlap the solid line in both subplots.

TABLE I. LOAD VOLTAGE DISLOCATION WHEN BRIDGE CURRENT DROPS TO ZERO

| Step Size | Method | | | | | |
|---|---|---|---|---|---|---|
| | A | | B | | C | |
| μs | Multi. | Dislo. | Multi. | Dislo. | Multi. | Dislo. | Multi. |
| 5 | -- | 1.47E-03 | -- | 1.83E-06 | -- | 1.83E-06 | -- |
| 10 | 2.0 | 2.93E-03 | 2.0 | 7.30E-06 | 4.0 | 7.30E-06 | 4.0 |
| 20 | 2.0 | 5.83E-03 | 2.0 | 2.91E-05 | 4.0 | 2.91E-05 | 4.0 |
| 40 | 2.0 | 1.16E-02 | 2.0 | 1.15E-04 | 4.0 | 1.15E-04 | 4.0 |
| 80 | 2.0 | 2.28E-02 | 2.0 | 4.52E-04 | 3.9 | 4.52E-04 | 3.9 |
| 160 | 2.0 | 4.45E-02 | 1.9 | 1.74E-03 | 3.9 | 1.74E-03 | 3.9 |
| 320 | 2.0 | 8.39E-02 | 1.9 | 6.43E-03 | 3.7 | 6.43E-03 | 3.7 |

Note: Dislo. stands for dislocation. Multi. stands for multiple, defined as the item on the left divided by the previous item. The same abbreviations are used in other tables in this paper.

and 60 Hz respectively. The capacitance is 0.02. The resistance load is 0.1. The firing angle for the thyristors is π/6 rad.

Fig. 3 compares the results from the time stepping scheme adopting a 100 μs step size with the three methods to the Simulink results. The feasibility of the three methods is demonstrated in that the time stepping scheme is able to obtain reasonably close results to Simulink with these methods. The spike in bridge current at about 0.03472 s is due to the direct connection between the voltage source and the capacitor, which causes an instantaneous change in capacitor voltage when the circuit is turned on by the firing signal. When bridge current drops to zero at about 0.03987 s, load voltage is continuous.

Fig. 3 shows that Method A is less accurate. Its results exhibit slight mismatch from the Simulink results. Load voltage from Method A experiences a visible dislocation when bridge current drops to zero. On the other hand, results from Methods B and C basically overlap the corresponding Simulink results, indicating their higher accuracy. Table I compares the load voltage dislocation of the three methods given different step sizes when bridge current drops to zero. Here, the dislocation is defined as the absolute value of the difference between load voltages at $t_{SW-}$ and $t_{SW+}$. Ideally, the dislocation should be zero because load voltage is physically continuous at the time instant.

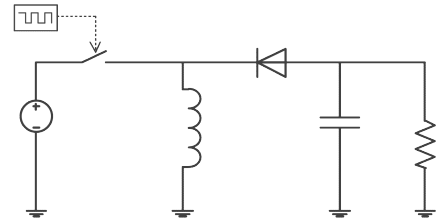

Figure 4. Buck-boost converter driving a resistance load.

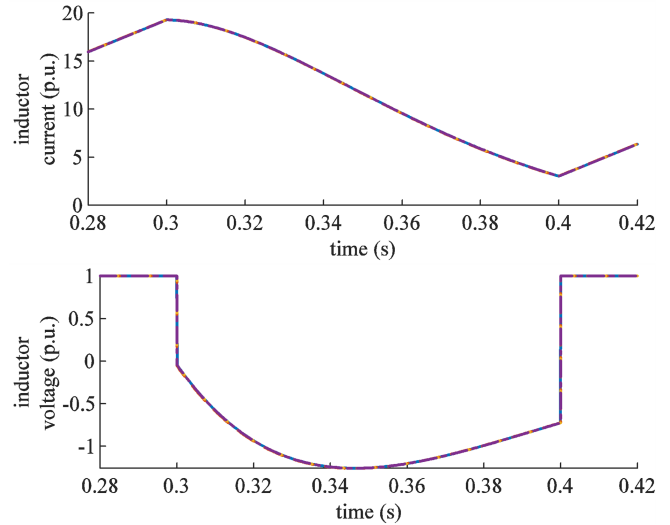

Figure 5. Results from the buck-boost converter driving a resistance load. Solid line: Simulink. Dashed line: Method A. Dotted line: Method B. Dash-dotted line: Method C. Jumps are precisely represented as vertical lines rather than sharp slopes. The dashed line, dotted line and dash-dotted line basically overlap the solid line in both subplots.

Table I shows that Methods B and C are indeed more accurate than Method A given the same step size. The difference in accuracy between Method B and Method C is negligible. For Method A, the dislocation basically doubles as the step size doubles, indicating that it is of $1^{st}$ order accuracy. For Methods B and C, the dislocation basically increases fourfold as the step size doubles, indicating that they are of $2^{nd}$ order accuracy.

*B. Buck-Boost Converter*

A buck-boost converter driving a resistance load is shown in Fig. 4. The value of the DC voltage source is 1.0. The inductance is 0.006. The capacitance is 0.3. The resistance load is 0.1. The fully controlled switch is driven by a square wave, of which the frequency is 5 Hz and the duty ratio is 0.5. The switch is on at the high signal level while it is off at the low signal level.

The results from the time stepping scheme adopting a 200 μs step size with the three methods are compared to the Simulink results in Fig. 5. It is observed that all the three methods reach a high degree of agreement with Simulink, which demonstrates their feasibility. The switch is turned off at 0.3 s while it is turned on at 0.4 s. Inductor current is continuous at both turning points. The inductor current dislocation of the three methods given different step sizes when the switch is turned off and turned on is listed in Tables II and III respectively. The dislocation is defined as the absolute value of the difference between inductor currents at $t_{SW-}$ and $t_{SW+}$, which should be zero ideally due to the aforementioned physical continuity.

TABLE II. INDUCTOR CURRENT DISLOCATION WHEN THE SWITCH IS TURNED OFF

| Step Size | Method | | | | | |
|---|---|---|---|---|---|---|
| | A | | B | | C | |
| µs | Multi. | Dislo. | Multi. | Dislo. | Multi. | Dislo. | Multi. |
| 5 | -- | 3.50E-05 | -- | 1.33E-07 | -- | 1.33E-07 | -- |
| 10 | 2.0 | 7.04E-05 | 2.0 | 5.30E-07 | 4.0 | 5.30E-07 | 4.0 |
| 20 | 2.0 | 1.42E-04 | 2.0 | 2.12E-06 | 4.0 | 2.12E-06 | 4.0 |
| 40 | 2.0 | 2.91E-04 | 2.0 | 8.47E-06 | 4.0 | 8.47E-06 | 4.0 |
| 80 | 2.0 | 6.07E-04 | 2.1 | 3.38E-05 | 4.0 | 3.38E-05 | 4.0 |
| 160 | 2.0 | 1.31E-03 | 2.2 | 1.35E-04 | 4.0 | 1.35E-04 | 4.0 |
| 320 | 2.0 | 3.02E-03 | 2.3 | 5.37E-04 | 4.0 | 5.37E-04 | 4.0 |

TABLE III. INDUCTOR CURRENT DISLOCATION WHEN THE SWITCH IS TURNED ON

| Step Size | Method | | | | | |
|---|---|---|---|---|---|---|
| | A | | B | | C | |
| µs | Multi. | Dislo. | Multi. | Dislo. | Multi. | Dislo. | Multi. |
| 5 | -- | 8.33E-04 | -- | 0 | -- | 0 | -- |
| 10 | 2.0 | 1.67E-03 | 2.0 | 0 | -- | 0 | -- |
| 20 | 2.0 | 3.33E-03 | 2.0 | 0 | -- | 0 | -- |
| 40 | 2.0 | 6.67E-03 | 2.0 | 0 | -- | 0 | -- |
| 80 | 2.0 | 1.33E-02 | 2.0 | 0 | -- | 0 | -- |
| 160 | 2.0 | 2.67E-02 | 2.0 | 0 | -- | 0 | -- |
| 320 | 2.0 | 5.33E-02 | 2.0 | 0 | -- | 0 | -- |

TABLE IV. INDUCTOR VOLTAGE DISLOCATION WHEN THE SWITCH IS TURNED OFF

| Step Size | Method | | | | | |
|---|---|---|---|---|---|---|
| | A | | B | | C | |
| µs | Multi. | Dislo. | Multi. | Dislo. | Multi. | Dislo. | Multi. |
| 5 | -- | 1.59E-04 | -- | 1.34E-08 | -- | 1.34E-08 | -- |
| 10 | 2.0 | 3.18E-04 | 2.0 | 5.36E-08 | 4.0 | 5.36E-08 | 4.0 |
| 20 | 2.0 | 6.36E-04 | 2.0 | 2.14E-07 | 4.0 | 2.14E-07 | 4.0 |
| 40 | 2.0 | 1.27E-03 | 2.0 | 8.57E-07 | 4.0 | 8.57E-07 | 4.0 |
| 80 | 2.0 | 2.54E-03 | 2.0 | 3.42E-06 | 4.0 | 3.42E-06 | 4.0 |
| 160 | 2.0 | 5.06E-03 | 2.0 | 1.37E-05 | 4.0 | 1.37E-05 | 4.0 |
| 320 | 2.0 | 1.01E-02 | 2.0 | 5.46E-05 | 4.0 | 5.46E-05 | 4.0 |

When the switch is turned off, inductor voltage exhibits a jump in value, which is physically a right-continuous waveform. In other words, the value of inductor voltage is the same at $t_{SW}$ and $t_{SW+}$. Table IV lists the inductor voltage dislocation of the three methods given different step sizes when the switch is turned off. The dislocation is defined as the absolute value of the difference between inductor voltages at $t_{SW}$ and $t_{SW+}$. In the ideal situation, the dislocation should be zero.

From Tables II-IV, it is learned that Methods B and C are obviously more accurate than Method A given the same step size. When the switch is turned on, Methods B and C can even calculate inductor current precisely despite the step size. Comparing Methods B and C, their accuracy is indistinguishable. Tables II-IV again indicate that Methods B and C are of $2^{nd}$ order accuracy, while Method A is merely of $1^{st}$ order. Such observation is consistent with that from the previous study case.

## V. CONCLUSION AND FUTURE WORK

Accurate calculation of switching events is possible without the common presumption of continuous differential state variables in EMT simulation. As a proof of concept, this paper proposes Methods B and C of $2^{nd}$ order accuracy for such calculation, demonstrating their feasibility and analyzing their accuracy. The $1^{st}$ order Method A is also feasible but less accurate.

Future research efforts may be directed towards theoretical analysis on the performance of the proposed methods. Exploring more accurate methods for calculation of switching events may also be of interest.